\documentclass[twocolumn,preprintnumbers,amsmath,amssymb,showpacs,superscriptaddress]{revtex4}

\usepackage{graphicx}
\usepackage{dcolumn}
\usepackage{bm}

\def\figin{\includegraphics[scale=0.4]}

\begin{document}


\title{Centrality and pseudorapidity dependence of charged hadron 
production at intermediate $p_{T}$ in Au + Au collisions 
at $\sqrt{s_{_{NN}}}=130$ GeV}

\affiliation{Argonne National Laboratory, Argonne, Illinois 60439}
\affiliation{University of Birmingham, Birmingham, United Kingdom}
\affiliation{Brookhaven National Laboratory, Upton, New York 11973}
\affiliation{California Institute of Technology, Pasedena, California 91125}
\affiliation{University of California, Berkeley, California 94720}
\affiliation{University of California, Davis, California 95616}
\affiliation{University of California, Los Angeles, California 90095}
\affiliation{Carnegie Mellon University, Pittsburgh, Pennsylvania 15213}
\affiliation{Creighton University, Omaha, Nebraska 68178}
\affiliation{Nuclear Physics Institute AS CR, 250 68 \v{R}e\v{z}/Prague, Czech Republic}
\affiliation{Laboratory for High Energy (JINR), Dubna, Russia}
\affiliation{Particle Physics Laboratory (JINR), Dubna, Russia}
\affiliation{University of Frankfurt, Frankfurt, Germany}
\affiliation{Insitute  of Physics, Bhubaneswar 751005, India}
\affiliation{Indian Institute of Technology, Mumbai, India}
\affiliation{Indiana University, Bloomington, Indiana 47408}
\affiliation{Institut de Recherches Subatomiques, Strasbourg, France}
\affiliation{University of Jammu, Jammu 180001, India}
\affiliation{Kent State University, Kent, Ohio 44242}
\affiliation{Lawrence Berkeley National Laboratory, Berkeley, California 94720}
\affiliation{Massachusetts Institute of Technology, Cambridge, MA 02139-4307}
\affiliation{Max-Planck-Institut f\"ur Physik, Munich, Germany}
\affiliation{Michigan State University, East Lansing, Michigan 48824}
\affiliation{Moscow Engineering Physics Institute, Moscow Russia}
\affiliation{City College of New York, New York City, New York 10031}
\affiliation{NIKHEF, Amsterdam, The Netherlands}
\affiliation{Ohio State University, Columbus, Ohio 43210}
\affiliation{Panjab University, Chandigarh 160014, India}
\affiliation{Pennsylvania State University, University Park, Pennsylvania 16802}
\affiliation{Institute of High Energy Physics, Protvino, Russia}
\affiliation{Purdue University, West Lafayette, Indiana 47907}
\affiliation{University of Rajasthan, Jaipur 302004, India}
\affiliation{Rice University, Houston, Texas 77251}
\affiliation{Universidade de Sao Paulo, Sao Paulo, Brazil}
\affiliation{University of Science \& Technology of China, Anhui 230027, China}
\affiliation{Shanghai Institute of Applied Physics, Shanghai 201800, China}
\affiliation{SUBATECH, Nantes, France}
\affiliation{Texas A\&M University, College Station, Texas 77843}
\affiliation{University of Texas, Austin, Texas 78712}
\affiliation{Tsinghua University, Beijing 100084, China}
\affiliation{Valparaiso University, Valparaiso, Indiana 46383}
\affiliation{Variable Energy Cyclotron Centre, Kolkata 700064, India}
\affiliation{Warsaw University of Technology, Warsaw, Poland}
\affiliation{University of Washington, Seattle, Washington 98195}
\affiliation{Wayne State University, Detroit, Michigan 48201}
\affiliation{Institute of Particle Physics, CCNU (HZNU), Wuhan 430079, China}
\affiliation{Yale University, New Haven, Connecticut 06520}
\affiliation{University of Zagreb, Zagreb, HR-10002, Croatia}

\author{J.~Adams}\affiliation{University of Birmingham, Birmingham, United Kingdom}
\author{M.M.~Aggarwal}\affiliation{Panjab University, Chandigarh 160014, India}
\author{Z.~Ahammed}\affiliation{Variable Energy Cyclotron Centre, Kolkata 700064, India}
\author{J.~Amonett}\affiliation{Kent State University, Kent, Ohio 44242}
\author{B.D.~Anderson}\affiliation{Kent State University, Kent, Ohio 44242}
\author{D.~Arkhipkin}\affiliation{Particle Physics Laboratory (JINR), Dubna, Russia}
\author{G.S.~Averichev}\affiliation{Laboratory for High Energy (JINR), Dubna, Russia}
\author{Y.~Bai}\affiliation{NIKHEF, Amsterdam, The Netherlands}
\author{J.~Balewski}\affiliation{Indiana University, Bloomington, Indiana 47408}
\author{O.~Barannikova}\affiliation{Purdue University, West Lafayette, Indiana 47907}
% ...

\collaboration{STAR Collaboration}\homepage{www.star.bnl.gov}\noaffiliation

\date{\today}

\begin{abstract}
We present STAR measurements of charged hadron production as a function of
centrality in Au + Au collisions at $\sqrt{s_{_{NN}}}=130$ GeV. 
The measurements cover
a phase space region of $0.2<p_{T}<6.0$ GeV/$c$ in transverse momentum 
and $-1<\eta<1$ in pseudorapidity.
Inclusive transverse momentum distributions of charged hadrons in 
the pseudorapidity region  $0.5<|\eta|<1$
 are reported and compared 
to our previously published results for $|\eta|<0.5$.
No significant difference is seen for inclusive $p_{T}$ 
distributions of charged hadrons in these
two pseudorapidity bins.
We measured $dN/d\eta$ distributions
and truncated mean $p_{T}$ in a region of $p_{T}>p_{T}^{cut}$, and studied 
the results in the framework of participant and binary
scaling. 
No clear evidence is observed for participant scaling of charged hadron
yield in the measured $p_{T}$ region. The relative importance of hard
scattering process is investigated through binary scaling fraction of
particle production.
\end{abstract}


\maketitle

\section{Introduction}

Quantum Chromodynamics (QCD) is considered to be the underlying theory of 
the strong interaction which governs hadron production in nuclear
collisions. The strong interaction is usually divided into soft
processes, which involve  small momentum transfer, and hard processes,
which can be calculated using perturbative QCD. The Relativistic Heavy
Ion Collider (RHIC) experiments at the Brookhaven National Laboratory 
investigate properties and evolution of
matter at high temperature and energy density. 
At RHIC energies, the hard processes become more evident in comparison
to previous heavy ion experiments and can be used
to probe the early state of the collision system. 
A high energy parton produced via hard scattering 
may lose energy in the hot/dense medium through gluon bremsstrahlung 
and multiple scatterings before 
hadronization~\cite{quenching1,quenching2}, 
leading to a suppression of high $p_{T}$ hadron production.
 The magnitude of the energy loss provides an indirect signature
of QGP formation. Since parton energy loss is directly proportional to 
gluon density, the energy loss would be much larger in a partonic medium 
than in hadronic matter~\cite{quenchsignature}.

Partonic energy loss can be investigated through comparison of hadron
yield as a function of $p_{T}$ between nucleus-nucleus collisions and
$p+p$ or $\bar{p}+p$ collisions.  In order to do so, scaling factors
which account for the nuclear geometry, the number of participant
nucleons, $N_{part}$, and the number of binary
nucleon-nucleon collisions, $N_{bin}$, are calculated
from theoretical models.  
Experimental results from RHIC, including our earlier analyses in the
pseudorapidity region $|\eta|<0.5$, 
have indicated a suppression of hadron production for $p_{T}>2$ GeV/$c$ in
central Au + Au collisions relative to $p+p$ and $\bar{p}+p$ 
collisions~\cite{ptsuppression2,star200,ptsuppression1}. 
This is in contrast to the SPS 
result from central Pb + Pb collisions at $\sqrt{s_{_{NN}}}=17$ GeV, 
which shows an excess of $\pi^{0}$ production for $2<p_{T}<4$
GeV/$c$~\cite{wa98,detail}. The RHIC measurements are striking
considering that known nuclear effects, like the Cronin effect~\cite{cronin} 
and radial flow~\cite{na49}, tend to enhance hadron yields at high
$p_{T}$. The RHIC results for high $p_{T}$ hadron suppression agree
qualitatively with calculations based on fragmentation models, which
attribute the high $p_{T}$ hadron suppression to medium induced parton
energy loss~\cite{loss}. 

Another known nuclear effect, nuclear shadowing, also modifies particle 
production at high $p_T$.  Calculations of this
effect~\cite{shadowing1} based on the 
EKS98 shadowing parametrization~\cite{EKS}
predicted it to be small in the $p_T$ 
and pseudorapidity region covered in this measurement.  However, another 
study~\cite{shadowing2} 
found a much larger shadowing effect for heavy nuclei at 
RHIC.  Therefore, a measurement of particle production as a function of 
$p_T$ and pseudorapidity may provide a constraint on the shadowing effect.

Partonic energy loss may also  
be studied by the pseudorapidity dependence
of hadron production.
Change of pseudorapidity due to change of momentum is
\begin{eqnarray}
\delta \eta= \frac{p_{z}}{p}(\frac{\delta p_{z}}{p_z}
	-\frac{\delta p_{T}}{p_T}).
\end{eqnarray}
The pseudorapidity distributions would be modified
as a result of the parton energy loss 
if the momentum change rate ($\delta p/p$) due to the energy loss is
different along the transverse and longitudinal direction. In
addition, Polleri and Yuan~\cite{yuan} pointed out that the degree of
the energy loss may also depend on  the pseudorapidity region in which
a jet is produced because the energy loss is proportional to the
particle density in pseudorapidity. The pseudorapidity dependence of
high $p_{T}$ hadron production provides a means to probe the initial
density of matter along both the transverse and longitudinal directions.

In this article, we present measurements of hadron production in Au +
Au collisions at $\sqrt{s_{_{NN}}}=130$ GeV as a function of
centrality, $p_{T}$ and $\eta$. In Sec.~II we will briefly describe
the STAR experimental setup and then give a description of data
analysis techniques that were used to obtain the inclusive transverse
momentum distributions for charged hadrons. We will also discuss the
parameterization of inclusive transverse momentum distributions in
$p+p$ collisions at $\sqrt{s}=130$ GeV and 
the calculations of $N_{part}$ and $N_{bin}$. In Sec.~III 
results from the data analysis will be reported and compared with
model calculations. 
Physics implications of our measurements are discussed in Sec.~IV.
And  we will then summarize our measurements in Sec.~V.

\section{Analysis}

\subsection{Experimental Setup and Data}

Measurements presented in this article are based on two data sets
of Au + Au collisions at $\sqrt{s_{_{NN}}}=130$ GeV, which were
recorded by the STAR detector at RHIC. A detailed description of the
STAR detector can be found elsewhere~\cite{star}. The two data sets
comprise minimum bias  and central collision triggered
events which correspond to approximately the most central 10\% of the Au +
Au geometric cross section. Charged particle tracks of an event were
detected in the Time Projection Chamber~\cite{tpc} (TPC) with a
pseudorapidity coverage $|\eta|<1.8$ and complete azimuthal symmetry.
The transverse momentum of a track is determined by fitting a circle
through the transverse coordinates of the primary event vertex and the
space points along the track in the TPC. The total momentum can be
calculated using this radius of curvature in a 0.25 T magnetic field
and the polar angle of the track. The procedure involves a three
dimensional fit using three coordinates of the primary vertex
determined from all of the tracks reconstructed in the TPC. The
primary vertex position along the beam direction, $z_{vtx}$, has a
wide spread with one standard deviation about 100 cm. To increase
detection efficiency of the tracks within $|\eta|<1$, we required the
events to have a primary vertex $|z_{vtx}| < 75$ cm. After the event
selection cuts, the minimum bias data set contained $\sim$ 181k events
and the central data set contained $\sim$ 365k events.

Centrality selection is based on the uncorrected primary charged particle
multiplicity $N_{ch}$ within $|\eta|<0.75$ and $p_{T}<1.5$
GeV/$c$. The requirement on $\eta$ range maximizes
the number of tracks used to define centrality in an event  
while keeping the tracking acceptance  approximately
constant. 
The percentage of the geometric cross section is determined in the
same way as that published by STAR previously~\cite{hminus}, where the
negatively charged hadron multiplicity $N_{h^-}$ distribution within
$|\eta|<0.5$ was used. The data set is divided into seven centrality
bins, and the most central bin is 0--5\% (the top 5\% of the
multiplicity distribution) while the most peripheral bin is 60--80\%. 

The analysis in this article covers a transverse momentum region 
of $0.2<p_{T}<6.0$ GeV/$c$. Accepted
primary tracks have $|\eta|<1$, at least 25 space points in the TPC used in
the track fit out of 45 pad rows,
a fit probability of being a primary track greater than
0.05, and a distance of closest approach to the primary vertex
less than 1 cm.   These track quality cuts were varied to estimate the
systematic uncertainty.
Acceptance and efficiency were determined by embedding simulated
tracks into actual Au + Au collision events. 

The measured high $p_{T}$ hadron yield is sensitive to small spatial
distortions of the TPC alignments in both azimuthal and longitudinal
directions. A measurement of the summed hadron yield, $(h^+ + h^-)/2$,
is less sensitive to such distortions than the yield of one charge
sign alone. Using 12 sectors from each of the TPC ends as
independent detectors for high $p_{T}$ hadrons, we estimated the
sectorwise (azimuthal direction) variations of the yields to be less
than 5\%. The variation of the yield between the hadrons crossing and
not crossing the
central membrane of the TPC was
found to be approximately proportional to $p_{T}$  with a value of 11\% at
$p_{T}=5.5$ GeV/$c$. The typical correction factors for the acceptance
and efficiency are given in Table~\ref{tab:errors} as
``Tracking''. The systematic uncertainties incorporate acceptance,
efficiency, track quality cuts, and the effects of the spatial
nonuniformity.  
The tracking and other correction factors and their systematic
uncertainties given in Table~\ref{tab:errors} for $|\eta|<0.5$ differ
from those given in our previous paper~\cite{ptsuppression2} because different
track quality cuts and 
other correction procedures were used.


Finite momentum resolution tends to spread particles to neighboring bins in a
momentum histogram, especially for an exponentially falling spectrum. 
This  smearing effect cannot be
neglected at higher $p_{T}$ where
the momentum resolution is limited by the strength of the magnetic
field and the TPC spatial resolution. We used
the embedding technique to determine 
the $p_{T}$ resolution.
For $p_{T}>0.5$ GeV/$c$ within $|\eta|<0.5$ the Gaussian distribution
of track curvature $k \propto 1/p_T$ has a relative width of 
$\delta k/k = 0.013 + 0.015 p_T/$(GeV/$c$) for central events and  
$\delta k/k = 0.012 + 0.012 p_T/$(GeV/$c$) for
peripheral events.
 Within $0.5<|\eta|<1$, 
$\delta k/k = 0.014 + 0.010 p_T/$(GeV/$c$) for central events and  
$\delta k/k = 0.014 + 0.0072 p_T/$(GeV/$c$) for
peripheral events.

The fact that the $p_{T}$ resolution for $0.5<|\eta|<1$ is better than that
for $|\eta|<0.5$ is due to the competition between two opposing effects.
For a given $p_{T}$ track in the TPC, the
hadron with higher $\eta$ tends to have fewer space points, hence
poorer resolution, but shorter
drift distance, hence better
resolution.

The magnitude of the $p_{T}$ resolution determined from the
embedding technique did not include the effect of the
primary vertex resolution. 
The effects of the
$p_{T}$ smearing due to the primary vertex resolution, 
to the charge-sign-dependent distortion,  
and to the weak decay background tracks, have been empirically derived from
the comparison between real and embedded tracks.
The combined effect 
within $0.5<|\eta|<1$ was
found to be larger than that within $|\eta|<0.5$. This 
is partially due to the fact that the magnitude of the 
charge-sign-dependent distortion in the higher $\eta$ region is larger.

The two contributions have been convoluted into a power law function
to fit the data, and then the ratio of the fitted function to its
convoluted one 
gives the $p_{T}$ smearing correction factor~\cite{YC}. 
Because the two contributions have opposite $|\eta|$ dependence,
the overall $p_{T}$ smearing correction factors for the two $\eta$ regions
happen to be comparable.
 The typical $p_{T}$ smearing correction factors and their systematic
uncertainties are also given in Table~\ref{tab:errors}.

\subsection{Background}

\begin{figure}
\figin{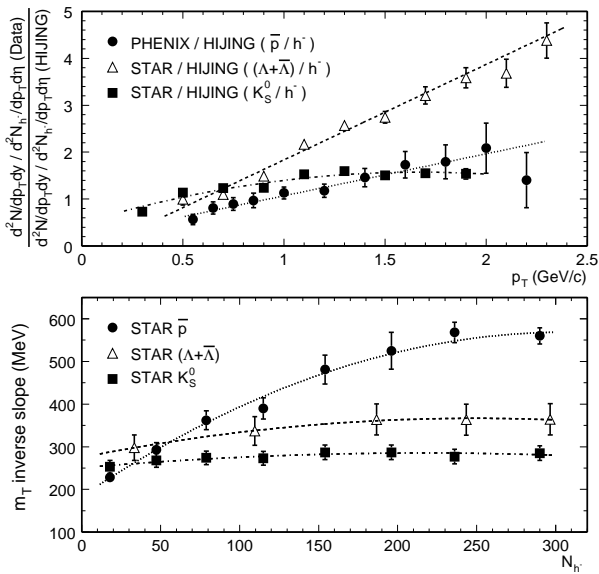}
\caption{\label{fig:bkg} 
Measurements used in background studies.
Upper panel: ratios of the measured $p_{T}$ yield
ratios to those of HIJING in the most central bin. Lower panel: measured
$m_{T}$ inverse slope parameters as functions of centrality
represented by measured negatively charged hadron multiplicity
$N_{h^-}$ within $|\eta|<0.5$.
Curves are polynomial fits to data points.
}
\end{figure}

The most significant backgrounds for the high $p_{T}$ charged hadron yield
as seen in Table~\ref{tab:errors}
come from particle weak decays and antinucleon annihilation in
detector material.
The contamination rate for each background source was
estimated using detector response simulations with events
generated by the HIJING model~\cite{hijing}. 
However, the $p_{T}$ dependence of production of weakly decaying particles,
primarily $K_S^0$, $\Lambda$, $\bar{\Lambda}$, and of antinucleons,
$\bar{p}$, $\bar{n}$, in HIJING is not consistent with experimental
measurements.  We corrected those predicted yields using the measured
spectra of $\bar{p}$~\cite{phenixpbar,starpbar}, $\Lambda$ and
$\bar{\Lambda}$~\cite{starlambda}, and $K_S^0$~\cite{starkshort}, together
with those of $h^-$~\cite{phenixpbar,hminus}, for $p_{T}<2.4$
GeV/$c$ in the mid-rapidity region in the most central bin. The corrections
used in calculating the background fractions are shown in the upper
panel of Fig.~\ref{fig:bkg}. The curves are polynomial fits to the
data points and are used in the interpolation due to different $p_{T}$
binning. For $p_{T}>2.4$ GeV/$c$ we simply assumed the yield ratios 
to be constant. Systematic
uncertainties of 50\% and 100\% of the overall background fraction are
assigned for the regions of $p_{T}<2.4$ GeV/$c$ and $p_{T}>2.4$ GeV/$c$,
respectively~\cite{YC}. 

The contamination rate for all background sources shows almost no
centrality dependence from the Monte Carlo HIJING events. Therefore,
centrality dependence of the background fraction is mainly determined
by the measured spectra in various centrality bins. 
In the lower panel of Fig.~\ref{fig:bkg}
we show the measured transverse mass ($m_T=\sqrt{p_{T}^2 + m_{0}^2}$) 
inverse slope parameters of exponential fits to
$\bar{p}$~\cite{starpbar},  
$\Lambda$ and
$\bar{\Lambda}$~\cite{starlambda}, and $K_S^0$~\cite{starkshort}
spectra in the mid-rapidity region as functions of the measured
negatively charged hadron multiplicity $N_{h^-}$ within
$|\eta|<0.5$. We use these to correct for different centrality binning
in our analysis. The polynomial fits are used to interpolate the
inverse slope parameters in the centrality bins used in this analysis. 

Pseudorapidity dependence of the background fraction
is studied using the Monte Carlo HIJING events. For $p_{T}<2$
GeV/$c$ the $\eta$-dependence of backgrounds is negligible within
$-1<\eta<1$ while for $p_{T}>2$ GeV/$c$ the background fraction decreases
with  $p_{T}$ and $|\eta|$. For example, at $p_{T}=5.5$ GeV/$c$
the background fraction predicted within $0.5<|\eta|<1$ is only 40\%
of that within $|\eta|<0.5$. The typical background correction factors
and their systematic uncertainties are given in Table~\ref{tab:errors}.
 The total systematic uncertainties of
the measured spectra within $|\eta|<0.5$ ($0.5<|\eta|<1$) at the highest bin
$p_{T}=5.5$ GeV/$c$ are $\approx 24\%$ ($\approx 18\%$) for central events
and $\approx 17\%$ ($\approx 15\%$) for peripheral events.

\begin{center}
\begin{table*}
\caption{ Typical multiplicative correction factors and systematic
uncertainties, applied to the yields for peripheral and central
collisions within $|\eta|<0.5$ and within $0.5<|\eta|<1$.
}
\begin{tabular}{|c|l|c|c|c|c|}
\hline
\hline
 Pseudorapidity  &  & \multicolumn{2}{c|}{ $p_{T}=2$ GeV/$c$ } &
\multicolumn{2}{c|}{ $p_{T}=5.5$ GeV/$c$ } \\ \cline{2-6}
 & Centrality & 60--80\% & 0--5\% 
& 60--80\% & 0--5\% \\ 
\hline
 & Tracking & $1.16\pm 0.10$ & $1.71\pm 0.15$ 
& $1.22\pm 0.16$ & $1.65\pm 0.22$ \\
$|\eta|<0.5$ & $p_{T}$ Smearing & $1.01\pm 0.01$ & $1.00\pm 0.01$ 
& $0.89\pm 0.02$ & $0.70\pm 0.06$ \\
 & Background & $0.92\pm 0.04$ & $0.88\pm 0.06$ 
& $0.90\pm 0.10$ & $0.85\pm 0.15$ \\ 
\hline
 & Tracking & $1.29\pm 0.11$ & $1.78\pm 0.15$ 
& $1.31\pm 0.18$ & $1.71\pm 0.23$ \\
$0.5<|\eta|<1$ & $p_{T}$ Smearing & $1.01\pm 0.01$ & $1.01\pm 0.01$ 
& $0.89\pm 0.02$ & $0.72\pm 0.07$ \\
 & Background & $0.92\pm 0.04$ & $0.88\pm 0.06$ 
& $0.96\pm 0.04$ & $0.94\pm 0.06$ \\ 
\hline 
\hline
\end{tabular}
\label{tab:errors}
\end{table*}
\end{center}

\subsection{$NN$ Reference}

In the absence of any $NN$ collision data at $\sqrt{s}=130$~GeV, a
$NN$ reference spectrum is obtained by extrapolation of the UA1
$\bar{p}+p$ data for  $\sqrt{s}= 200-900$~GeV~\cite{ua1}. 
The UA1 inclusive charged particle $p_{T}$ spectra within $|\eta|<2.5$ 
were fitted by the pQCD inspired power law
function 
\begin{equation}
\frac{1}{2\pi p_{T}}\frac{d^{2}N}{dp_{T}d\eta} = 
C \left( 1+ \frac{p_{T}}{p_0} \right) ^{-n}.
\label{eq:power}
\end{equation}
The fit parameters were used to extrapolate to our energy, giving
$C\sigma_{in} = 267^{+4}_{-6}$~mb/(GeV/$c$)$^2$ ($\sigma_{in}$
denotes the inelastic cross section of $NN$ collisions),
$p_0=1.90^{+0.17}_{-0.09}$~GeV/$c$, and $n=12.98^{+0.92}_{-0.47}$ at 
$\sqrt{s}=130$~GeV~\cite{ptsuppression2}.
The superscripts and subscripts are curves that bound the systematic
uncertainty. 

\begin{figure}
\figin{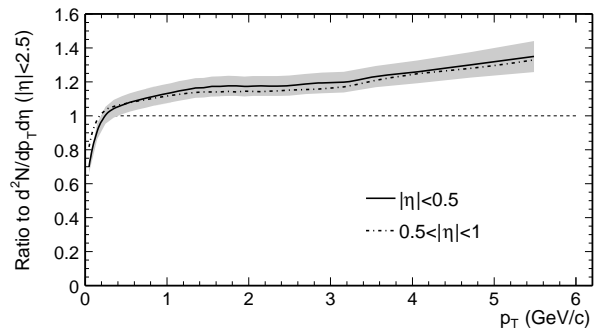}
\caption{\label{fig:nn} $\eta$ acceptance correction function from
model calculations: ratios of $p_{T}$ spectra in two different $\eta$
regions to that within $|\eta|<2.5$, in which the UA1 Collaboration
published its inclusive charged particle $p_{T}$ spectra. }
\end{figure}

However, the UA1 acceptance is different from STAR's. 
Corrections were made to the UA1 reference for our $\eta$ acceptance based on
two independent pQCD calculations: PYTHIA~\cite{pythia} and
Vitev~\cite{vitev}.
When the $K$ factor in PYTHIA is set to 1.5, PYTHIA calculations for 200 GeV
$\bar{p}+p$ collisions
are in reasonable agreement 
with  the UA1 measurement of the inclusive charged particle $p_{T}$
spectrum~\cite{ua1}  
and with the UA5 measurement of the pseudorapidity density
distribution~\cite{ua5}. Similar PYTHIA calculations 
are in reasonable agreement 
with the STAR measurement of the inclusive charged
hadron $p_T$ spectrum within $|\eta|<0.5$ for $p+p$ collisions at
$\sqrt{s}=200$ GeV~\cite{star200}.
Fig.~\ref{fig:nn} shows the $p_{T}$-dependent correction functions for
two $\eta$ regions at $\sqrt{s} = 130$ GeV, 
obtained by averaging over the two pQCD calculations. 
The solid curve is the ratio of $d^{2}N/dp_{T}d\eta$ within
$|\eta|<0.5$ to that within $|\eta|<2.5$, and the shaded area shows its
systematic uncertainty. The dot-dashed curve shows the same ratio for
$0.5<|\eta|<1$, and the similar magnitude of the uncertainty 
on the ratio of $0.5<|\eta|<1$ to $|\eta|<2.5$ 
is not shown. Multiplicative corrections of $1.35\pm
0.09$ and $1.33\pm 0.09$ at $p_{T}=5.5$ GeV/$c$ have been obtained for
$|\eta|<0.5$ and for $0.5<|\eta|<1$, respectively.
The difference between $|\eta|<0.5$ and $0.5<|\eta|<1$ is quite small,
indicating a relatively flat $\eta$ distribution within $-1<\eta<1$
for a broad $p_{T}$ region. 
The STAR measurement~\cite{star200} 
is consistent with the UA1 $\bar{p}+p$ data for 200 GeV 
after applying a similar $\eta$ acceptance correction.

We derived $\sigma_{in}$ in the $NN$ reference at $\sqrt{s}=130$~GeV 
of $40 \pm 3$ mb by requiring $dN/d\eta(|\eta|<0.5)$, 
which was obtained by integrating the extrapolated spectrum after applying
the $\eta$ acceptance correction, to be
2.25, which was determined from the energy dependence 
of $dN/d\eta(\eta=0)$~\cite{cdfeta}. 

\subsection{Participant and Binary Collision Determination}

The number of participant nucleons, $N_{part}$, and the number of
binary nucleon-nucleon collisions, $N_{bin}$, in a nucleus-nucleus
collision are used to compare experimental results with model
predictions. Unfortunately, at RHIC  $N_{part}$ and  $N_{bin}$ cannot
be measured directly and have to be obtained in a model-dependent
way. Considerable discrepancy exists among various model calculations,
especially for peripheral collisions~\cite{sa}. 

We first investigate  $N_{part}$ and $N_{bin}$ obtained from a Monte
Carlo (MC) Glauber model calculation~\cite{ptsuppression2,starpi}. 
In the Monte Carlo Glauber model, each of the nucleons in a nucleus A is
randomly distributed using a
Woods-Saxon nuclear density distribution
\begin{equation}
\rho(r) = \frac{\rho_{0}}{1+\exp[(r-r_{0})/ D]},
\label{eq:ws}
\end{equation}
with normalization to $\int{\rho(r)dr} = {\rm A}$ and parameters: 
nuclear radius $r_{0}$ and surface diffuseness $D$. 
All nucleons in either nucleus for a nucleus-nucleus collision are
required to be separated by a minimum distance. 
The calculated $d\sigma/dN_{part}$ or $d\sigma/dN_{bin}$
distribution was divided into bins corresponding to common
fractions of the total geometric cross section to extract the average
$N_{part}$ or $N_{bin}$ for each centrality bin. The systematic
uncertainties on $N_{part}$ and $N_{bin}$ were estimated by varying the
Woods-Saxon parameters, by varying the $\sigma_{in}$ value, and by
including a 5\% uncertainty in the determination of the total geometric
cross section. 

\begin{center}
\begin{table*}
\caption{Comparisons of nuclear geometries implemented in various
models for 130 GeV Au + Au collisions. }
\begin{tabular}{|c|c|c|c|}
\hline
\hline
Model  &HIJING 1.35 &VENUS 4.12 & MC Glauber\\
\hline
Woods-Saxon& $r_{0}=6.38$ fm & $r_{0}=6.64$ fm &
$r_{0}=6.5\pm 0.1$ fm \\
Parameters  &  $D=0.535$ fm &  $D=0.540$ fm &  $D=0.535\pm 0.027$ fm \\
\hline 
Minimum Distance &  & & \\
of Two Nucleons  & 0.4 fm & 0.8 fm & 0.4 fm\\
\hline 
 Nucleon-Nucleon 
  &        $\Omega(b)=(1+\sigma_{jet}/\sigma_{soft})\chi_{0}(\xi) $
& & \\
Overlap Function& 
$\xi=b/b_{0}(s) $ & $\theta(R-b)$ & $\theta(R-b)$ \\
    & $\chi_{0}(\xi)=\mu_{0}^2(\mu_{0}\xi)^3 K_{3}(\mu_{0}\xi)/96$
& & \\
\hline 
Maximum Impact &  &  & \\
Parameter & 25.6 fm & 24.1 fm & no restriction \\
\hline
Nucleon-Nucleon &  & & \\
Cross Section $\sigma_{in}$  &  38.7 mb & 37.4 mb &
$41\pm 1$ mb\\
\hline
Total Geometric & & & \\
Cross Section & 7.27 b & 7.34 b & $6.9\pm 0.4$ b \\
\hline 
\hline
\end{tabular}
\label{tab:models}
\end{table*}
\end{center}

We also investigate calculations of $N_{part}$ and $N_{bin}$ using two
dynamic  models, HIJING~\cite{hijing} and VENUS~\cite{venus}. 
We compare these calculations with results from the Monte Carlo
Glauber model calculation to shed light on 
the model dependent uncertainties of $N_{part}$ and $N_{bin}$.

The VENUS model is based on the Gribov-Regge theory and string  
fragmentation.
The HIJING generator is an example of a two-component
model:
the momentum transfer of the soft process is treated 
phenomenologically and the hard processes are calculated by
pQCD. The excited nucleons after
collisions are stretched out as quark-diquark strings 
and fragment based on the Lund fragmentation
scheme~\cite{lund}. The parton energy loss in dense medium (quenching)
and nuclear modification
of parton structure functions (shadowing) are also modeled in HIJING. 

Both dynamic models describe nuclear
collision geometry using 
the Woods-Saxon nuclear density distribution and the eikonal formalism 
to determine the probability for each binary nucleon-nucleon
collision, and to compute $N_{part}$ and $N_{bin}$.
Table~\ref{tab:models} shows the
comparisons of the nuclear geometries implemented in HIJING, 
VENUS, and the Monte Carlo Glauber model for Au + Au collisions at
$\sqrt{s_{_{NN}}}=130$ GeV. 
The overlap
function, which defines the probability for a nucleon-nucleon collision 
at a given impact parameter $b$, has the form of $1-\exp(-2\Omega(b))$ 
in  HIJING 
with $\Omega(b)$ defined in Table~\ref{tab:models} 
($\mu_{0}=3.9$ and $\pi b_{0}^2(s)=\sigma_{soft}(s)/2$) while it is a step
function, $\theta(R-b)$, 
in VENUS and MC Glauber. 

The correspondence
between the centrality classes defined 
by measured charged particle multiplicity
and those defined by modeled impact parameter was used to extract
the average $N_{part}$ and $N_{bin}$ from these dynamic models for a
given centrality bin. 
Variations of average $N_{part}$ and $N_{bin}$ for different centrality 
selections were estimated 
using the Monte Carlo
events from the HIJING model.
The event classes corresponding to the same fractional cross section 
were selected by cuts on
$b$, $N_{ch}$, $N_{part}$, and $N_{bin}$. 
The average $N_{part}$ and  $N_{bin}$ by different cuts in HIJING
are consistent within 2\% for each centrality bin except 
the 60--80\% most peripheral bin, where the discrepancy is at a
level of 6\%.

\begin{figure}
\figin{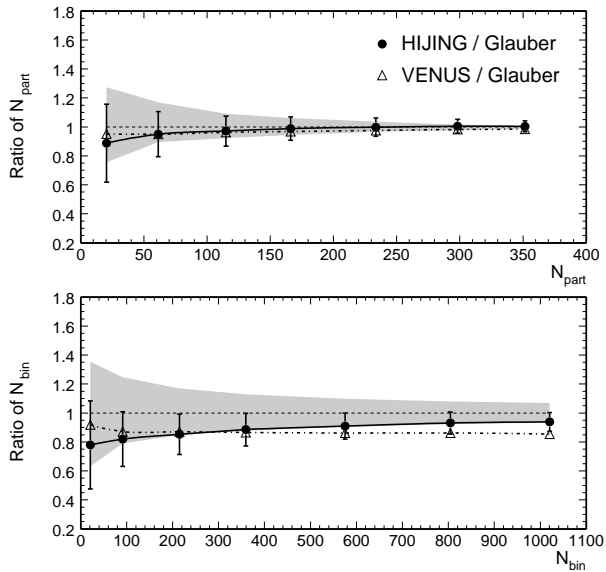}
\caption{\label{fig:npart} Ratio of the number of participants
$N_{part}$ (upper panel) or the number of binary collisions $N_{bin}$
(lower panel) 
determined from different models to that from a Monte Carlo Glauber 
calculation. 
Shaded areas
show the uncertainties of $N_{part}$ or $N_{bin}$ from 
the Monte Carlo Glauber calculation.
Curves are to guide the eye.
}
\end{figure}

The results of $N_{part}$ and $N_{bin}$ from the models are shown in 
Fig.~\ref{fig:npart} as their ratios to $N_{part}$ or $N_{bin}$
from the Monte Carlo Glauber calculation.
The participant scaling exponents $\alpha$, which are 
defined in the expression of 
$ N_{bin} = B \cdot N_{part}^{\;\alpha}$,
were obtained by fit to be $1.41 \pm 0.08$, $1.34 \pm 0.08$,  and
$1.38 \pm 0.08$ for HIJING, VENUS, and MC Glauber. 
The scaling exponents $\alpha$ for these models are
approximately  $4/3$ due to the fact that
$N_{part}\propto {\rm A}^1$ and $N_{bin}\propto {\rm A}^{4/3}$.

It is worthwhile to note here that distribution differences among
HIJING and VENUS are mainly due to different overlap functions.
The $N_{part}$ and $N_{bin}$ distributions 
from them are nearly identical
if the same overlap functions are used in these two model calculations.
Fig.~\ref{fig:npart} shows that over a broad range of centrality the
model dependent uncertainties of $N_{part}$ and $N_{bin}$ are within 10\% and
20\%, respectively.

\section{Results}

\begin{figure}
\figin{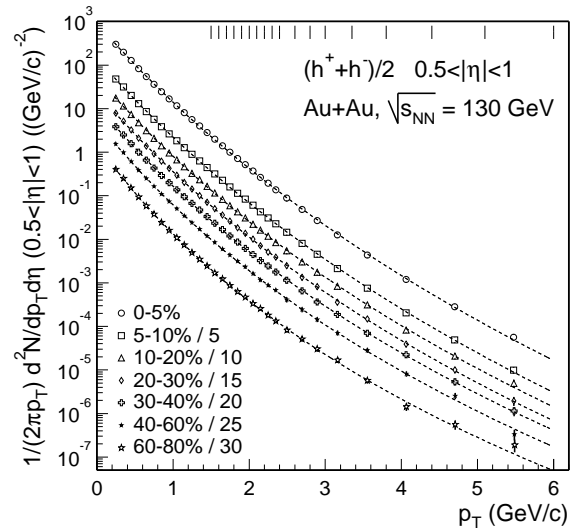}
\caption{\label{fig:spec} Inclusive $p_{T}$ distributions of $(h^+ +
h^-)/2$ within $0.5<|\eta|<1$. Noncentral bins are scaled down by the
indicated factors. The combined statistical and systematic errors are
shown. Curves are fits to the power law function. Hash marks at the
top indicate bin boundaries for $p_{T}>1.5$ GeV/$c$.
}
\end{figure}

Inclusive $p_{T}$ distributions of $(h^+ +
h^-)/2$ within $|\eta|<0.5$ have been published
previously~\cite{ptsuppression2}. The independent analysis reported in this
article shows that the differences to the published results for all
measured $p_{T}$ points are within 10\%,
 which is comparable to the systematic uncertainties for $p_{T}<2$ GeV/$c$ and
is less than the systematic uncertainties for the high $p_{T}$ region.
Fig.~\ref{fig:spec} shows inclusive $p_{T}$ distributions of $(h^+ +
h^-)/2$ within $0.5<|\eta|<1$ for various centrality bins. 
The error bars are the quadrature sum of statistical error and systematic
uncertainty, and are dominated by the latter except for the highest
$p_{T}$ point in the peripheral bins. The curves in Fig.~\ref{fig:spec} are 
power law function (Eq.~\ref{eq:power}) fits to the spectra.

\begin{figure}
\figin{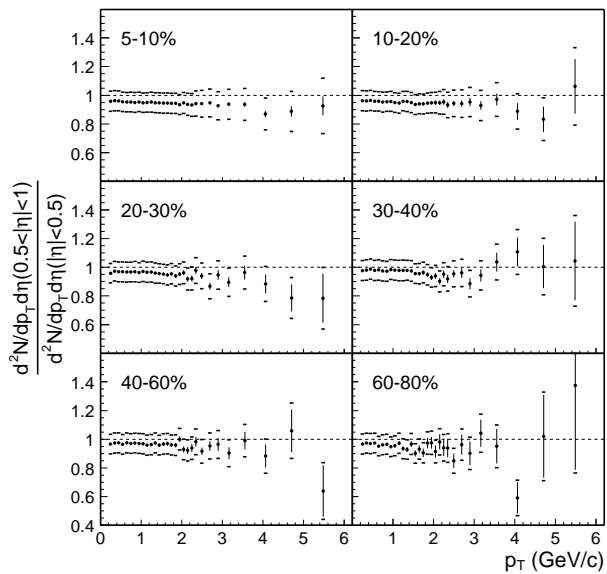}
\caption{\label{fig:ratio} Ratios of $p_{T}$ distributions within
$0.5<|\eta|<1$ to those within $|\eta|<0.5$ in various centrality
bins. Error bars show statistical errors while caps are the quadrature sum of
statistical errors and remaining systematic uncertainties.
}
\end{figure}

Fig.~\ref{fig:ratio} shows ratios of $p_{T}$ distributions within
$0.5<|\eta|<1$ to those within $|\eta|<0.5$ in various centrality bins.
Note that Fig.~\ref{fig:ratio} and the succeeding figures 
utilize the $p_T$ distributions 
within $|\eta|<0.5$ obtained here.  Using identical cuts and correction 
procedures across the full pseudorapidity region minimizes the systematic 
uncertainties in the relative comparisons.
 The error bars in Fig.~\ref{fig:ratio} show
statistical errors only while the caps are the quadrature sum of
statistical errors and systematic uncertainties
 which cannot be canceled out. Remaining systematic uncertainty
includes the variation due to track quality cuts, the uncertainties of
the $p_{T}$
smearing corrections for the two $\eta$ regions, and the partial uncertainty
of background subtraction related to the $\eta$-dependent part discussed
in Sec.~II.  

\begin{figure}
\figin{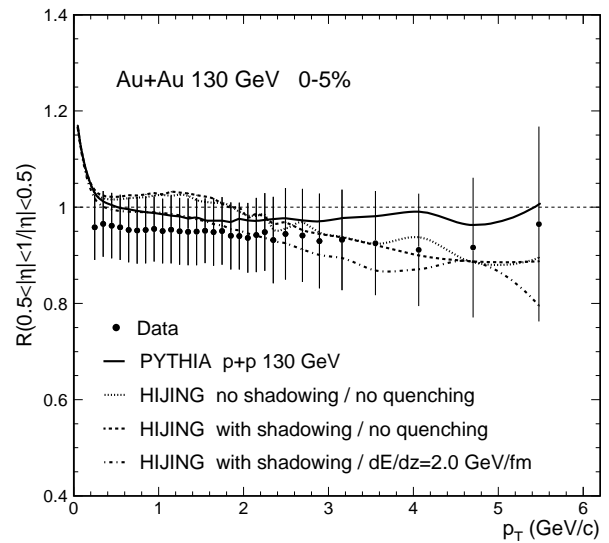}
\caption{\label{fig:cent} Ratio of $p_{T}$ distribution within
$0.5<|\eta|<1$ to that within $|\eta|<0.5$ in the  0--5\% most central
bin. Points are measurements and error bars include statistical and
remaining systematic uncertainties.  Curves are described in the text.
}
\end{figure}

Fig.~\ref{fig:cent} shows the same ratio of $0.5<|\eta|<1$ to $|\eta|<0.5$ 
in the 0--5\% most central bin. The points are our
measurements and the error bars include statistical and remaining
systematic uncertainties.  The solid curve is the same ratio from
PYTHIA calculations~\cite{pythia} for 130 GeV $p+p$ collisions.  
Other curves are ratios from HIJING
predictions of 130 GeV Au + Au collisions 
without shadowing and without quenching (dotted curve), with shadowing 
and without quenching (dashed curve), and with shadowing
and with partonic energy loss being 2.0 GeV/fm (dot-dashed curve).
The results show that the effects on the pseudorapidity dependence of both 
nuclear shadowing and partonic energy loss 
as implemented in HIJING  are too small to be tested in the measured
kinematic region under current experimental uncertainties.

No significant differences are observed in the comparisons of the inclusive
charged hadron yields between the two $\eta$ regions in
Fig.~\ref{fig:ratio} and Fig.~\ref{fig:cent} 
over a broad range of centrality for
all measured $p_{T}$ points. It suggests that an approximate boost
invariant condition might be established in the early stage of
collisions. 
The suppression pattern of the particle yield has little $\eta$
dependence in the measured region though the particle yield itself is
sensitive to partonic energy loss. 
A measurement of this ratio between 
$\eta=2.2$
and $\eta=0$ from the BRAHMS Collaboration shows that the ratio 
is below unity 
at $p_{T} \sim 4$ GeV/$c$~\cite{brams}.


\begin{figure}
\figin{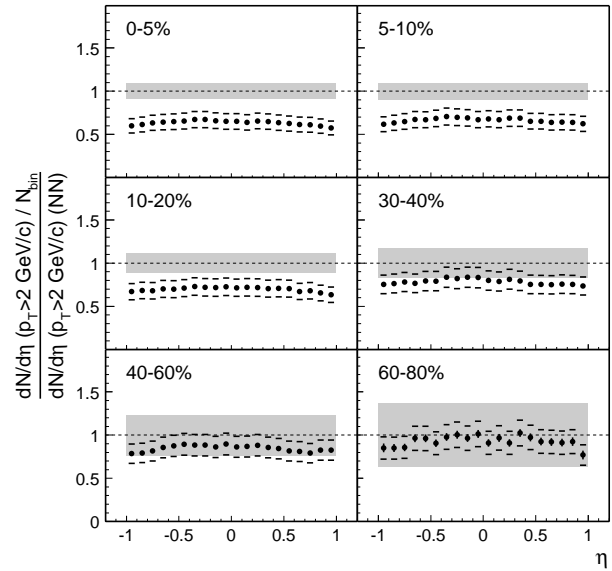}
\caption{\label{fig:pt2} $dN/d\eta$ distributions for $p_{T}>2$ GeV/$c$
and $-1<\eta<1$ scaled by $N_{bin}$ and divided by the $NN$ reference. 
}
\end{figure}

Fig.~\ref{fig:pt2} shows $dN/d\eta$ distributions for $p_{T}>2$ GeV/$c$ and 
$-1<\eta<1$ in various centrality bins. The error bars show
statistical uncertainties while the caps are the quadrature sum of
statistical and systematic uncertainties. The systematic uncertainties
are dominant and highly correlated. The $dN/d\eta$ distributions are scaled by
$N_{bin}$ 
and divided by the $NN$ reference. Due to nearly
complete $\eta$-independence of the $NN$ reference data 
for $p_{T}>0.2$ GeV/$c$ within $-1<\eta<1$   
as shown in
Fig.~\ref{fig:nn} and Fig.~\ref{fig:cent}, constant $dN/d\eta$
 of the $NN$ reference are used in Fig.~\ref{fig:pt2}.
Therefore, the shapes of the $dN/d\eta$ distributions for the Au + Au
collisions are preserved. 
The 
uncertainties on both $N_{bin}$ and the $NN$ reference data are shown
in the shaded regions around the lines at unity which represent the binary
collision scaling. Ratios below unity in the figure show that the high $p_{T}$
hadrons over 2 GeV/$c$ are suppressed with respect to those in $p+p$
collisions. 
The shape of $dN/d\eta$ for the high $p_{T}$ hadrons is nearly flat.
No significant centrality dependence of the $dN/d\eta$ shapes within
$-1<\eta<1$ is observed.
Similar behaviors are observed for $p_T>4$ GeV/$c$
except larger
suppressions in the central bins. For example, in the 0--5\%
most central bin, the average ratio is $0.41\pm 0.10$ for $p_T>4$
GeV/$c$ 
while it is $0.64\pm 0.10$ for $p_T>2$ GeV/$c$.

\section{Discussion}


The charged hadron yield per participant pair 
at $\sqrt{s_{_{NN}}}=130$ GeV shows a slow increase as a function of
$N_{part}$~\cite{phob,firstphenix,ptsuppression2}. Such 
slow increase of hadron multiplicity as a function of centrality at
RHIC  has been considered by Kharzeev {\sl et al}.~\cite{satur2} in
the framework of parton saturation. They argued that the hadron
multiplicity as a function of centrality would increase faster if produced jets
lose energy radiating soft gluons that in turn fragment into hadrons
at mid-rapidity.  As a result of the parton saturation, 
it is predicted that hadron multiplicity should scale with  $N_{part}$
at a moderately high $p_{T}$ (up to 6--8 GeV/$c$ at RHIC energies).
An explanation of the slower than expected increase in fragmentation
models is that the
effective energy loss is significantly reduced in a thermal
environment due to detailed balance~\cite{detail}. 
Recent experimental results in  
$d$ + Au collisions at $\sqrt{s_{_{NN}}}=200$ GeV support the idea
that the suppression of high $p_{T}$ hadron production in Au + Au collisions 
at mid-rapidity is due to final state interactions
rather than parton saturation in the initial state~\cite{brams,dau}. 

\begin{figure}
\figin{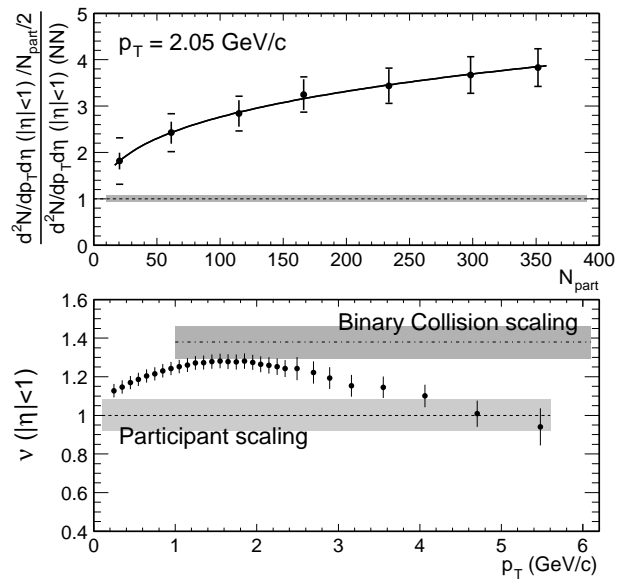}
\caption{\label{fig:dnsc} 
Upper panel: ratio of charged hadron yields 
within $|\eta|<1$
for Au + Au relative to the $NN$
reference, scaled by $N_{part}/2$ as a function of centrality 
for a  $p_{T}$ bin at $p_{T}= 2.05$ GeV/$c$. 
The curve is fit to $B \cdot N_{part}^{\nu}$.
Lower panel:
 participant scaling exponent $\nu$
of charged hadron yields as a function of $p_{T}$ 
within $|\eta|<1$.
}
\end{figure}

In the upper panel of Fig.~\ref{fig:dnsc}, we plot charged hadron
yield per participant pair within $|\eta|<1$ normalized to 
that of nucleon-nucleon collisions as a  function of  $N_{part}$ for
$p_{T}=2.05$ GeV/c. 
The error bars are the
 uncertainties of data while the caps are the quadrature sum of
the uncertainties of both data and $N_{part}$.
The shaded regions around unity show systematic uncertainties of the $NN$ 
reference data. 
The result shows that the ratio is above unity and increases with $N_{part}$. 

Dependence of the  charged hadron yield 
on $N_{part}$  
can be studied by fitting the yield by the following function
\begin{equation}
 \frac{d^2N}{dp_Td\eta} = B\cdot N_{part}^{\nu}
\end{equation}
in different $p_T$ bins. Such an example is 
shown as a curve in the upper panel of Fig.~\ref{fig:dnsc} for
$p_T = 2.05$ GeV/$c$.
The fit parameter, $\nu(p_{T})$, 
is given in  the lower panel of Fig.~\ref{fig:dnsc} 
as a function of $p_{T}$. 
 The error bars are
the uncertainties of the fit parameters associated with 
the uncertainties of data.
The lines and shaded regions are
binary collision ($N_{bin}$) and participant
($N_{part}$) scaling exponents and uncertainties to 
$N_{part}$. No clear evidence of participant scaling over the whole 
measured $p_{T}$ region is observed. The approximate participant scaling 
of the hadron yield at high 
$p_{T}$ observed by PHOBOS~\cite{phobos} 
appears to be accidental.

\begin{figure}
\figin{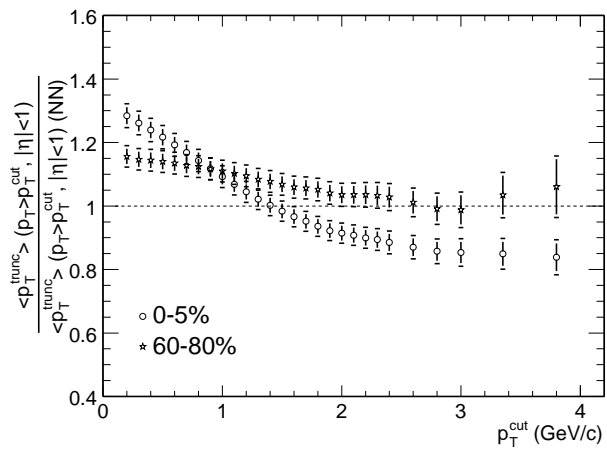}
\caption{\label{fig:par}
Ratio of truncated mean $p_{T}$ in
$p_{T}>p_{T}^{cut}$
within $|\eta|<1$
as a function of $p_{T}^{cut}$ for central and peripheral collisions.
}
\end{figure}

In a scenario with continuous energy loss of particles through a
medium, the energy loss would lead to a shift in the mean $p_{T}$ of
these particles. If the energy loss contributes to additional particle
production in the low $p_{T}$ region, the mean $p_{T}$ of low $p_{T}$
particles would also be modified.
The truncated mean $p_{T}$, defined as
\begin{eqnarray}
\langle p_{T}^{trunc} \rangle (p_{T}^{cut}) = 
\frac{\int_{p_{T}^{cut}}^{\infty}p_{T}\cdot dN/dp_{T}\cdot 
dp_{T}}{\int_{p_{T}^{cut}}^{\infty}dN/dp_{T}\cdot dp_{T}} - p_{T}^{cut},
\label{eq:avept}
\end{eqnarray}
is used to study the variation of mean $p_{T}$ as a function of
$p_{T}$ scale with respect to $NN$ reference data. Fig.~\ref{fig:par}
shows the truncated mean $p_{T}$ ratios between Au + Au and $p+p$
collisions as a function of $p_{T}^{cut}$ for central (0--5\%) and
peripheral (60--80\%) collisions.  The errors are combined statistical and
systematic uncertainties while the caps are the quadrature sum of the
uncertainties of both the Au + Au data and the $NN$ reference data. 

In peripheral collisions at high $p_{T}$ ($p_{T}^{cut} \agt 3$
GeV/$c$) the truncated mean $p_{T}$ of particles is approximately the
same as for $p+p$ collisions above the same $p_{T}^{cut}$
(Fig.~\ref{fig:par}). The ratio in the low $p_{T}$ region is above
unity indicating the effects of the Cronin effect and/or radial flow
in peripheral collisions. For central collisions, the truncated mean
$p_{T}$ for $p_{T}^{cut} \agt 3$ GeV/$c$ is approximately 15\% lower
than the truncated mean $p_{T}$ from $p+p$ collisions at the same
$p_{T}^{cut}$, consistent with the scenario for partonic energy loss
in this $p_{T}$ region. The significantly larger ratio in the 
low $p_{T}$ region probably reflects the combined effects of larger radial 
flow, the Cronin effect, and  $p_{T}$ shift  of particles due to
 energy loss, 
which cannot be decoupled with the present data.

%

\begin{figure}
\figin{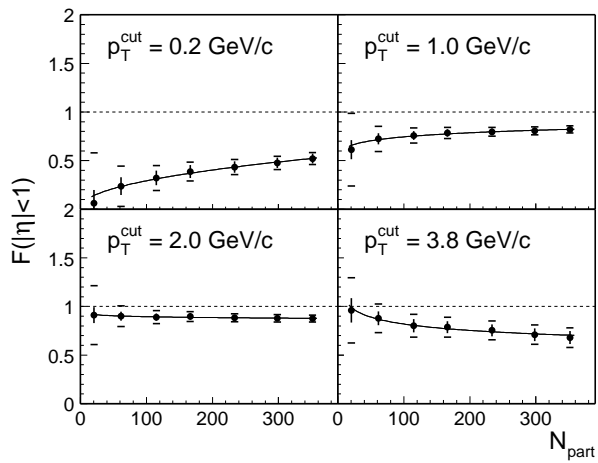}
\caption{\label{fig:hardcent}
Binary scaling fraction in $p_{T}> p_{T}^{cut}$ 
within $|\eta|<1$
as a function of centrality for selected $p_{T}^{cut}$. 
For $p_{T}^{cut}>2$ GeV/$c$,
the fraction $F$ decreases with centrality.
}
\end{figure}

Fig.~\ref{fig:dnsc} (bottom panel) indicates that over a broad
$p_{T}$ region particle production falls between participant and
binary scalings. In two-component particle production models, the
binary scaling has been associated with hard parton scatterings and
the participant scaling with the soft processes. In our study we
empirically decompose the particle yield into $N_{part}$ and $N_{bin}$
scaling components, i.e.
\begin{eqnarray}
dN/d\eta(p_{T}^{cut}) & = & 
(1-x(p_{T}^{cut})) \; n_{pp}(p_{T}^{cut}) \cdot \frac{N_{part}}{2}   
\nonumber \\
 &  & + \; x(p_{T}^{cut}) \; n_{pp}(p_{T}^{cut}) \cdot N_{bin},
\label{eq2}
\end{eqnarray}
where $n_{pp}(p_{T}^{cut})$ and $x(p_{T}^{cut})$ are the hadron
multiplicity and  the fraction of particle yield attributable to hard
processes in  $p+p$ collisions, respectively. Fig.~\ref{fig:hardcent}
shows the binary scaling fraction in  Au + Au collisions, defined as
\begin{eqnarray}
F(p_{T}^{cut}) = \frac{x(p_{T}^{cut})\; n_{pp}(p_{T}^{cut}) 
\cdot N_{bin}}{dN/d\eta(p_{T}^{cut})}.
\label{eq3}
\end{eqnarray}
Note that $F(p_{T}^{cut})$ does not depend on $n_{pp}(p_{T}^{cut})$
since both numerator and denominator of Eq.~\ref{eq3} contain
$n_{pp}(p_{T}^{cut})$. There is a distinguishable trend as a function
of $N_{part}$ from $p_{T}^{cut} = 3.8$ GeV/$c$ to lower $p_{T}^{cut}$. 
This trend is consistent with the $\nu(p_{T})$ dependence 
in Fig.~\ref{fig:dnsc}.
It is worth noting that $F\approx 70$\% in central Au + Au collisions
at $p_{T}^{cut}=3.8$ GeV/$c$. However, one should exercise
caution when relating this fraction to hard parton scattering
processes, particularly at lower $p_{T}$ where high $p_{T}$ particles
may suffer large energy losses in the medium and become soft.

\section{Conclusion}



We have presented inclusive distributions of $(h^+ + h^-)/2$ from STAR at 
RHIC in the region $0.5<|\eta|<1$ and compared them to distributions  
for $|\eta|<0.5$, finding no significant differences in the region of
$0.2<p_{T}<6.0$ GeV/$c$. We find that the $dN/d\eta$ distributions for
$-1<\eta<1$ are nearly flat for all centralities. The charged hadron
yield as a  function of $p_{T}$ shows no clear participant scaling in
the measured $p_{T}$ region. The binary scaling fraction in the
two-component model shows a decrease with centrality for
$p_{T}^{cut}>2$ GeV/$c$ and is about 70\% at $p_{T}^{cut}=3.8$ GeV/$c$
for central collisions.

\begin{acknowledgments}

We thank the RHIC Operations Group and RCF at BNL, and the
NERSC Center at LBNL for their support. This work was supported
in part by the HENP Divisions of the Office of Science of the U.S.
DOE; the U.S. NSF; the BMBF of Germany; IN2P3, RA, RPL, and
EMN of France; EPSRC of the United Kingdom; FAPESP of Brazil;
the Russian Ministry of Science and Technology; the Ministry of
Education and the NNSFC of China; SFOM of the Czech Republic,
FOM and UU of the Netherlands,
DAE, DST, and CSIR of the Government of India; the Swiss NSF.

\end{acknowledgments}

{}

\end{document}